\documentclass[fleqn,5p,twocolumn]{elsarticle}
\usepackage{science_pkg}
\usepackage{multicol}
\usepackage{setspace}
\begin{document}

\begin{frontmatter}
\title{Inter-sensor propagation delay estimation using sources of opportunity}

\author[adcea,adgipsa]{Rémy Vincent} 
\author[adcea]{Mikael Carmona}
\author[adgipsa]{Olivier Michel}
\author[adcea,adgipsa]{Jean-Louis Lacoume}

\address[adcea]{CEA, Leti, 38054 Grenoble, France}
\address[adgipsa]{GIPSA-lab, BP 46 F-38402 Grenoble Cedex, France}

\begin{abstract}
Propagation delays are intensively used for Structural Health Monitoring or Sensor Network Localization. In this paper, we study the performances of acoustic propagation delay estimation between two sensors, using sources of opportunity only. Such sources are defined as being uncontrolled by the user (activation time, location, spectral content in time and space), thus preventing the direct estimation with classical active approaches, such as TDOA, RSSI and AOA. Observation models are extended from the literature to account for the spectral characteristics of the sources in this \textit{passive} context and we show how time-filtered sources of opportunity impact the retrieval of the propagation delay between two sensors. A geometrical analogy is then proposed that leads to a lower bound on the variance of the propagation delay estimation that accounts for both the temporal and the spatial properties of the sources field.
\end{abstract}
\end{frontmatter}

\setcounter{tocdepth}{3}


\section{Introduction to passive estimation} \label{sec:intro}

Propagation delay estimation is a fundamental issue for several applications such as Structural Health Monitoring (SHM) and Sensor Network Localization (SNL). For instance, structural monitoring can be achieved by relating structural deformations to the variations of a propagation delay, between two sensors of known location. SNL can be achieved from a set of inter-sensor propagation delays by using dedicated algorithms (see \textit{e.g.} \cite{Patwari}) and benefits SHM for at least two reasons; firstly it allows to associate the measurements (e.g. temperature) to the precise sensor locations, which can be crucial for diagnosis; secondly, it enables the retrieval of physical or geometrical parameters from the geometry of the sensor network itself (\textit{e.g.} velocity maps in seismology \cite{Olabode}).

Recently, SNL in aerial acoustics was used for an SHM application, namely to retrieve the bending of a beam, see \cite{Vincent}. Microphones were distributed at the surface of the structure, and identification of the shape of the beam was conducted by processing acoustic pressure fields propagating around the beam, unlike mechanical displacements.

\begin{figure}[ht!]
   	 \centering
	 \includegraphics[scale=0.5]{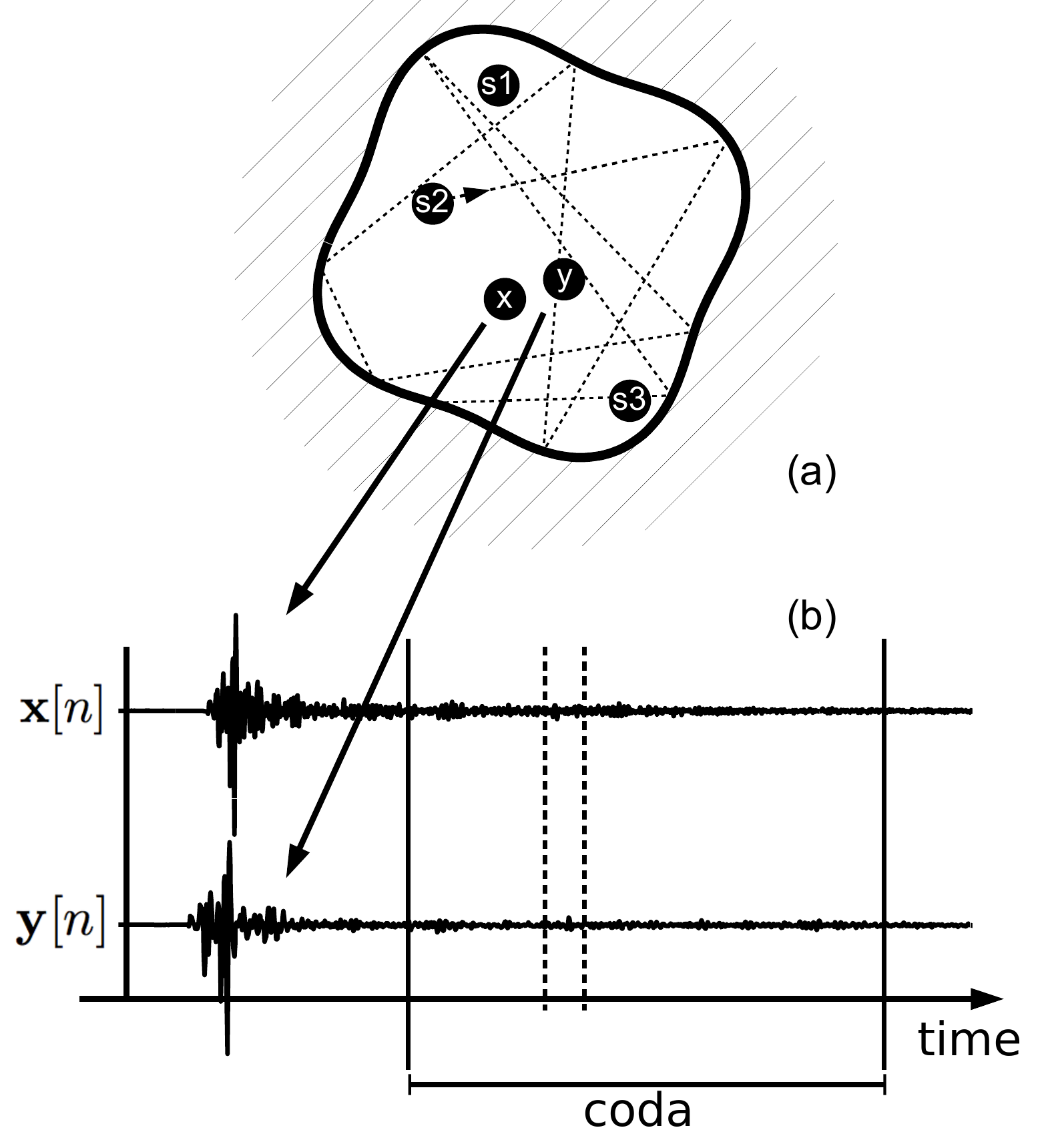}
	 \caption{(a) Schematic of the propagation medium. Sources s1, s2 and s3 are sources of opportunity. Their location and spectral content are uncontrolled. (b) Response of a source of opportunity (a handclap).}
	 \label{fig:signals}
\end{figure}

Furthermore, as in \cite{Sabra}, SNL was performed using sources of opportunity only. These latter are defined as sources that are uncontrolled (unknown positions, activation and spectral content), thus preventing the direct estimation of the inter-sensor distance with classical active approaches, such as TDOA, RSSI and AOA, see \textit{e.g.} \cite{Zhao,Leung,Dumont,Chalise,Blandin}. Fig. \ref{fig:signals} displays an example of source of opportunity in aerial acoustic: a handclap. In this case, the tail of the recorded responses are spatially correlated, due to the propagation in the medium. In section \ref{s:ip}, a technique is presented to retrieve the Green's function between the two sensing locations from those responses. Exploiting sources of opportunity to estimate parameters of a propagation medium is usually referred to as passive identification \cite{Carmona}, output-only identification \cite{Farrar} or seismic interferometry \cite{Campillo,Wapenaar}. Unlike active methods, passive methods do not require any additional equipment to the sensor network, nor does it require the structure to be temporarily put out-of-order.

\hfill

Pioneering work in passive identification from Nyquist \cite{Nyquist} is related to the fluctuation dissipation theorem \cite{Kubo} and establishes the relationship between the two-point correlation of the field and the Green’s function between those two locations, provided the source field is white in time and space. Major contributions can be found in seismic interferometry for the passive imaging of the Earth interior \cite{Campillo, Campillo2, Aki, Snieder, Gouedard,Prieto,Sanchez}, acoustics (ultrasounds) \cite{Lobkis, Sabra} and electromagnetism \cite{Wapenaar} for the local imaging of heterogeneities. Early contributions to SHM can be found under the name output-only structural analysis, with \cite{Farrar} illustrates the passive monitoring of a bridge from its ambient mechanical vibrations. More recent work can be found alongside with formalization efforts, see \textit{e.g.} \cite{ColinDeVerdiere,Snieder,Siringoringo}.

Interestingly, the passive identification protocols have proven to be robust and repeatable in various media and at various wavelengths. However, the propagation velocities of mechanical displacement fields prevents to localize at the scale of the structure, because of the current technological limitations. In this regard, acoustic wave propagation was shown in \cite{Vincent} to lead to a millimetric precision for the inter-sensors distance estimation. Thus, acoustic propagation (in the bandwidth $[0.1 - 20]$ kHz) represents a complementary approach for passive structural analysis. Nonetheless, compared to the low frequency and narrow band signals from long range propagation considered in seismic applications, pressure fields in aerial acoustics are rather wide-band and high frequency. As a consequence, there is a need to evaluate the performances of passive estimators in aerial acoustics and to take the evolutions of dissipation effects over the considered frequency range into account. To the best of our knowledge, such performances study was never derived in a systematic way in the literature. The closest investigations to our purpose were those of \cite{Sthely}, in which the impact of instrumentation time shifts on travel-time retrieval is studied. The effects of the spatial distribution of sources on the retrieval of the Green's function for a free acoustic propagation medium were observed and formalized in \cite{Derode,Roux,Sanchez}, but did not lead to performances assessment.

In this paper, we propose an estimation theoretic approach for passive distances retrieval in the context of homogeneous acoustic propagation. The first contribution is the derivation of a passive estimator of inter-sensor distances, or delays of propagation, that  generalizes existing equations obtained in an ideal framework (when the opportunity sources are white in time and space), see \textit{e.g.} \cite{Weaver,ColinDeVerdiere}. Secondly, a geometrical analogy of the passive estimation process is proposed which leads to analytical developments. From the proposed observation model, a Cramer-Rao lower bound is derived for the propagation delay estimation using sources of opportunity only. Simulations and real-data experiments illustrate the relevance of our results and their applications in a practical context.
	
This paper is organized as follows:
\begin{itemize}
	\item The passive context is first presented and the identity that provides an estimator of Green's functions is extended for fields that are not white in time;
	\item Three passive estimators are then investigated, for the Green's correlation function, for the Green's function and for the propagation delay. They are studied in terms of the operations required to compute them and the parameters that influence their performances;
	\item A novel observation model is then derived that allows to account for both the spatial and temporal properties of the sources, in the passive estimation process;
	\item A lower bound on the mean square error for the passive propagation delay estimation is presented.
\end{itemize}
\section{Passive Green's function retrieval} \label{s:ip} 

Based on the prior work of \cite{ColinDeVerdiere}, three passive estimators are presented that account for non whiteness (in time) of the source over the considered bandwidth; estimators for the Green's correlation, the Green's function and the propagation delay.

\subsection{Propagation equation and Green's function} 

Let $\mc{X}$ be a linear acoustic wave propagation medium, also assumed isotropic and homogeneous. The propagation equation that describes fields in fluids and in linear media with damping was derived by Stokes, see \textit{e.g.} \cite{Stokes,Buckingham}. The pressure field $f$ is related to the source field $s$ by
\begin{equation}
\left[\frac{\partial^2}{\partial t^2} + \eta \Delta \frac{\partial}{\partial t} + v^2\Delta \right]f(t,\mb{x}) = s(t,\mb{x}).
\label{m:Au=f} 
\end{equation}
where $v$ is the wave velocity and $\eta$ is the kinematic viscosity, related to damping. Indeed, the above equation displays a damping term $\eta \Delta \frac{\partial}{\partial t} f(t,\mb{x})$, which indicates that damping increases proportionally to the cube of the frequency. As short to medium range propagation distances are considered --- between source and sensors, as well as between two sensors --- high frequency components of the source spectra are significant and cannot be overlooked.

When the propagation operator in Equation \eqref{m:Au=f} is invertible, its kernel is the Green's function $g(t-t_0,\mb{x},\mb{y})$. It is the impulse response in $\mc{X}$ between locations $\mb{x}$ and $\mb{y}$, as the source transmits at time $t_0$. The Green's function allows to relate sources and responses by a filtering process generalized to time and space variables. For unbounded media, it reads
\begin{equation}
f(t,\mb{x}) = \iint_{\mbb{R}^3\times\mbb{R}} g(t-t',\mb{x}-\mb{u})s(t',\mb{u})dt'd\mb{u}.
\label{m:u=Gf}
\end{equation}

In the following, the convolution operator $*$ is used to describe filtering operations. Subscripts indicate the dimension (time/space) along which the convolution is performed, such that Equation \eqref{m:u=Gf} is then equivalent to $f(t,\mb{x})=[g \underset{t,\mb{u}}{*} s](t,\mb{x})$.  

\subsection{Green's correlation and Ward identity} \label{ssec:gc} 

Pioneering work in \cite{Weaver} elaborates on the fluctuation dissipation theorem and shows in acoustics that there exists a relationship between the two-point correlation of the field and the Green's function between those two locations, provided the source field is white in time and space. This identity is here extended to sources in aerial acoustics (in the audio range $[0.1 - 20]$ kHz) that are not white in time. Firstly, the correlation of the field $f$ recorded at $\mb{x}$ and $\mb{y}$ is defined as
\begin{eqnarray}
\gamma_{f}(u,v,\mb{x},\mb{y}) = \mathbb{E}\big[f(u,\mb{x})f^*(v,\mb{y}) \big],
\label{m:cu}
\end{eqnarray}
where the expectation operator $\mbb{E}$ applies on both the time and the space variable. 

In the literature, see for instance \cite{Weaver,Lobkis,ColinDeVerdiere,Lacoume}, the source field is assumed to be stationary and white in both time and spatial domains. Its correlation function is
\begin{equation}
\mbb{E}\big[ s(t,\mb{x}) s^*(t+u,\mb{y}) \big] =  \delta(u) \delta( \mb{x} - \mb{y} ).
\label{m:corrsourceblanche}
\end{equation}

In this case, the correlation of the fields recorded is called the Green's correlation function. It is noted $\gamma_g$ and is obtained combining Equations \eqref{m:u=Gf}, \eqref{m:cu} and \eqref{m:corrsourceblanche},
\begin{equation}
\gamma_g(u,\mb{x},\mb{y}) = \big[ g \underset{t,\mb{u}}{*}  g^- \big](u,\mb{x}-\mb{y}).
\label{m:cg}
\end{equation}
where function $g^-$ is the time-reversed version of $g$, \textit{e.g.} $g^-(t,\mb{x},\mb{y}) = g(-t,\mb{x},\mb{y})$.

In \cite{Weaver,Gouedard,ColinDeVerdiere}, an identity is derived from Equation \eqref{m:cg} that relates the Green's correlation $\gamma_g$ function to the odd part of the Green's function $\texttt{odd}\big[ g \big] = \frac{1}{2}(g - g^-)$. The resulting relationship is called Ward identity and accounts for constant damping, which means the damping term in Equation \eqref{m:Au=f} is proportional to $\frac{\partial}{\partial t} f(t,\mb{x})$, such that
\begin{equation*}
\frac{\partial }{\partial t} \gamma_g (u,\mb{x}-\mb{y}) = -\frac{1}{\eta} \texttt{odd}\big[ g \big] (u,\mb{x},\mb{y}).
\label{m:wi}
\end{equation*}
where it appears that the damping ($\eta \neq 0)$ is necessary to establish the Ward identity. A Ward identity was more specifically derived for weakly viscous acoustic media in \cite{Carmona}. Weak viscosity allows to assume $\eta / v \lambda \ll 1$, where $\lambda$ is the wavelength. Under this assumption, the Ward identity can be approximated by
\begin{equation}
\frac{\partial^3 }{\partial t^3} \gamma_g (u,\mb{x}-\mb{y}) \approx \frac{1}{\eta} \texttt{odd}\big[ g \big] (u,\mb{x},\mb{y}).
\label{m:wi_wviscous}
\end{equation}

In other words, the Ward identity provides an estimator $\hat g(u,\mb{x},\mb{y})$, for the Green's function between locations $\mb{x}$ and $\mb{y}$. Because the Green's function is causal, retrieving its odd part suffices to retrieve it entirely,
\begin{equation}
\hat g(u, \mb{x},\mb{y}) = \frac{\partial^3 }{\partial t^3} \gamma_g (u,\mb{x}-\mb{y}), \textrm{ for } u\geq0.
\label{m:gest}
\end{equation}

\subsection{Generalization to non temporally white sources}

Unlike in Equation \eqref{m:cg}, assume instead that the (spatially extended) source $s$ is white in space but is stationary in time. Its time auto-correlation function is $\gamma_s(t)$ and its total correlation function is given by
\begin{equation}
\mbb{E}\big[ s(t,\mb{x}) s^*(t+u,\mb{y}) \big] =  \gamma_s(u) \delta( \mb{x} - \mb{y} ).
\label{m:corrs}
\end{equation}

Combining Equations \eqref{m:u=Gf}, \eqref{m:cu} and \eqref{m:corrs} results in an extended interferences formula
\begin{equation}
\mbb{E}\big[ f(t,\mb{x}) f^*(t+u,\mb{y}) \big] = \big[ \gamma_s \underset{t}{*} g \underset{t,s}{*}  g^- \big](u,\mb{x}-\mb{y}).
\label{m:Cf}
\end{equation}
where the correlation of the source field acts as a filter on the Green's correlation from Equation \eqref{m:cg}. Combining Equations \eqref{m:Cf} and \eqref{m:wi_wviscous} to the Green's function estimator $\hat g (u,\mb{x},\mb{y})$ in Equation \eqref{m:gest} results in an estimator that accounts for the spectral properties of the source field,
\begin{equation}
\hat g (u,\mb{x},\mb{y}) \approx \frac{1}{\eta} \texttt{odd}\big[ \gamma_s \underset{t}{*} g \big](u,\mb{x},\mb{y}).
\label{m:wig}
\end{equation}

As can be seen in Equation \eqref{m:wig}, sources of opportunity allow to retrieve Green's functions filtered by the correlation of the source, which is related to the spectral content of the source itself. For a source that is white in time and space, the Green's function could theoretically be retrieved with an arbitrarily small estimation residue, depending on the observation duration and the sensors noise only.
\section{Passive parameter estimation} \label{s:pest}

\subsection{Observation model} 

Physical parameters characterizing the propagation medium as presented in the introduction may be inferred  from the Green's function. Here, highlight is made on the passive retrieval of inter-sensor propagation delays. The minimum propagation delay is defined as the time required for a wave to travel from a position $\mb{x}$ to a position $\mb{y}$ by the shortest path. If the velocity of the wave in the medium is known, an inter-sensor distance can be easily retrieved from this minimum propagation delay.

Assume that the dimensions of the medium $\mc{X}$ are much larger than the wavelength; then quasi-optical ray propagation is considered \cite{Schuster}. The Green's function can be modeled as,
\begin{equation}
g(t,\mb{x},\mb{y}) = \sum_{i=1}^\infty a_i \delta(t-t_i), \textrm{ with } |a_i| < |a_j|,~i < j,
\label{m:gmodel_full}
\end{equation}
where the amplitudes decay due to damping and geometrical attenuation. In the following, detection/estimation performances are presented for the passive minimum propagation delay estimation, $t_1$. From model \eqref{m:gmodel_full}, a simple estimator $\hat t_1$ for $t_1$ can be
\begin{equation}
\hat t_1  =  \arg\max_t ~\lbrace g( t,\mb{x}, \mb{y}) \rbrace.
\label{m:test}
\end{equation}

In the passive context, it was shown in Equation \eqref{m:wig} that the Green's function is estimated as a function of the source auto-correlation function. If the Green's function is retrieved at short times, it emulates the response of the same medium but with no bounds, thus involving the first propagation delay only. In this case, an observation model for the retrieved Green's function can be
\begin{equation}
\hat g(t) = A\bar\gamma_s(t-t_1) + w(t),
\label{m:gmodel}
\end{equation}
where $w$ is an additive estimation noise, assumed to be white Gaussian, with variance $\sigma^2$, for the sake of argumentation. $\bar\gamma_s$ is the normalized correlation of the source.

\subsection{Theoretical performances} \label{ss:theoretical_perf} 

\subsubsection{Detection} \label{sss:misdetection} 

In the active context, propagation delay estimation can be performed by locating the maximum amplitude of the inter-correlation between the reference source signal and the propagated signal. Interestingly, the observation model \eqref{m:gmodel} is very similar to the one encountered in an active context. In the passive context however, the source is unknown but is expected to be as white as possible ($\gamma_s \approx \delta$), to lead to optimal performances. For this reason, the propagation delay estimator \eqref{m:test} is thus justified in the passive context as well.

\begin{figure}[!ht] \centering
	\includegraphics[scale=0.4]{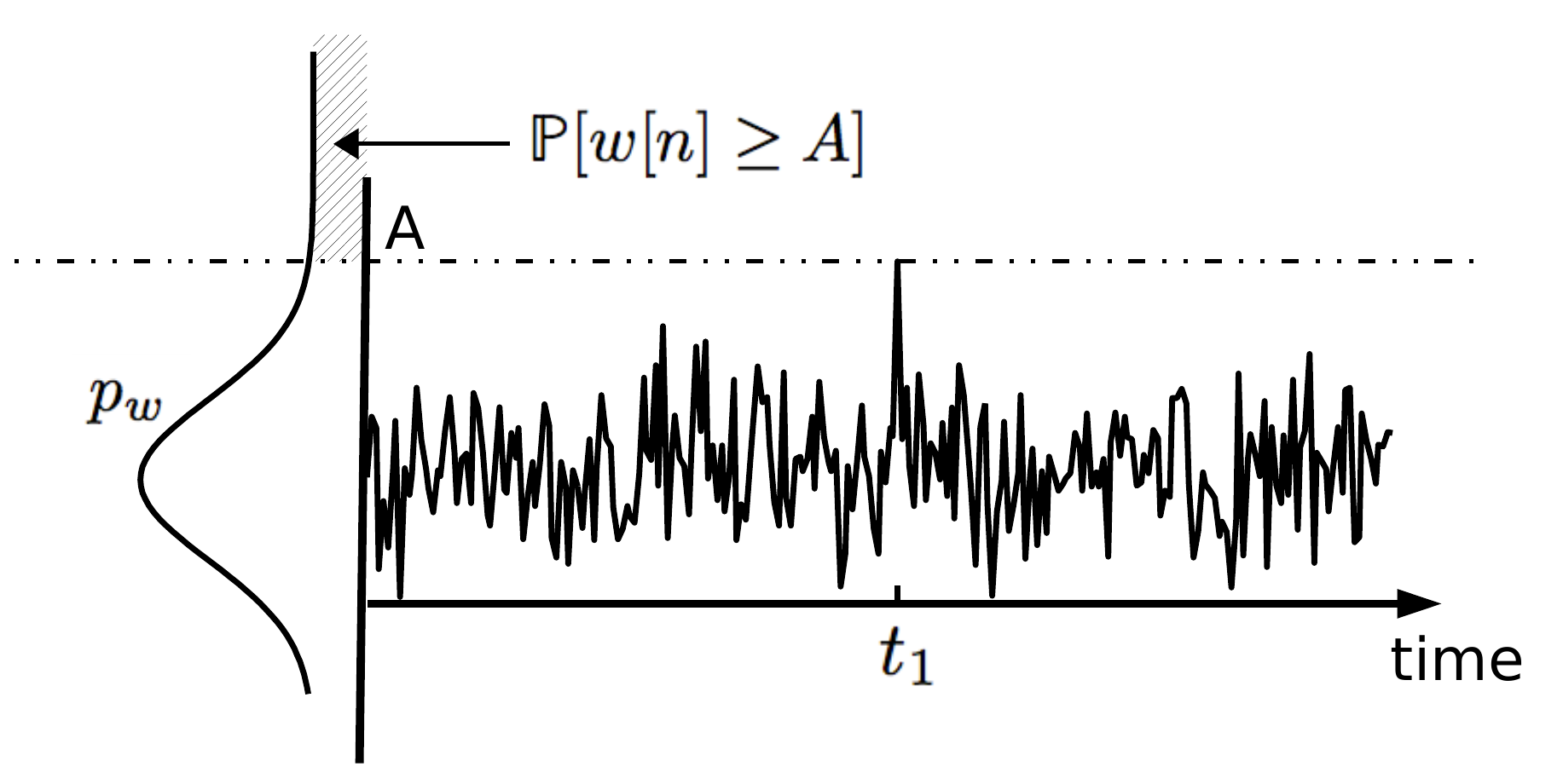} 
	\caption{Model for Green's function $\hat g(t) = A\delta(t-t_1) + w(t)$ and probability error of extracting the pulse location.}
	\label{f:pe}
\end{figure}

From model \eqref{m:gmodel}, taking the maximum amplitude of $\hat g(t)$ results in $A$, unless the amplitude of the estimation noise gets higher than $A$ at some point. The probability of this happening is a standard probability detection error, illustrated in Fig. \ref{f:pe}, see \cite{Ziv} for an example. The probability that the amplitude of the noise (assumed Gaussian) gets higher than $A$ writes
\begin{equation}
\mbb{P}[w[n] \geq A] = \frac{1}{2}(1-\mathrm{erf}(\frac{\sqrt{2}}{2}\frac{A}{\sigma})),
\label{m:perr}
\end{equation}
where $\mathrm{erf}(x) = 2/\sqrt{\pi} \int_0^x\exp(-t^2)dt$ is the error function. The probability increases as the ratio $A/\sigma$ decreases. Because the noise samples are assumed i.i.d., the probability of misdetection is proportional to the length of the signal. 

Recall that in the passive context, the Green's function is estimated using sources of opportunity. Thus, the minimum propagation delay in Equation \eqref{m:test} is extracted from the Green's function estimated using sources of opportunity. This estimation process is thus once again affected by the spectral content of the source field, as expressed in Equations \eqref{m:wig}. In the following, the content of the time frequency spectrum and the spatial frequency spectrum of the source are in turn discussed, with respect to their impact on passive estimation performances.

\subsubsection{Estimation} \label{sss:crlb} 

Let the source be white over the bandwidth B. The passive retrieval of the Green's function at short times (involving the first propagation delay only) is thus derived from model \eqref{m:gmodel} as
\begin{equation}
\hat g(t) = A~\mathrm{sinc}\big( B(t-t_1) \big) + w(t)
\label{m:gmodel2}
\end{equation}
where the estimation residue $w(t)$ is modeled Gaussian, with power spectral density constant equal to $N_w$ on the bandwidth $B$. Let the sampling rate be $\Delta = 1/(2B_s)$. Interestingly, this model applies for both the active and the passive context, tough with different levels of the signal of interest and of the noise. For this model, the lower bounds on the variance for the propagation delay estimation can be found in the literature, for instance in \cite{Kay1993}, such that
\begin{equation}
\mbb{E}\big[ (\hat t_1-t_1)^2\big] \geq \frac{N_w}{\frac{4}{3} \pi^2 A^2 B_s B^2 } 
\label{m:pcrlb1}
\end{equation}
where $\frac{A^2B}{4N_w}$ is shown to be a SNR in \cite{VanTrees} and $\frac{16}{3} \pi^2 B B_s$ is identified as the mean square bandwidth of the signal, such as in \cite{Ziv} for instance.


\subsection{Illustration for time-filtered signals} \label{ss:exp} 

In this section, the previous theoretical performances are illustrated on a real-data experiment, which consists of the passive estimation of the acoustic propagation delay between two microphones in a bounded environment. Lowpass filters with various cut-off frequencies were applied to the recorded signals and the estimation performances were computed for each of these band-limited signals.

Sources of opportunity can be pulses emitted at uncontrolled times, locations and with uncontrolled energies. In this case, it was shown in \cite{Campillo, Weaver} that the tail of those impulse responses, called coda waves \cite{Aki}, are the part of the signals that can be used to estimate the Green's function between the two sensing locations. Here, handclaps randomly distributed in the room, were used as sources of opportunity. In the experiment, two omni-directional microphones were placed $48$~cm away from each other, in a room with dimensions 10m$\times$6m$\times$3.5m.  Sensors recorded time series $\mb{x}[n]$ and $\mb{y}[n]$ at sampling frequency $44.1$~kHz, see Figure \ref{fig:signals}. As in \cite{Carmona}, the mixing time was used to set the beginning of the coda waves. A threshold was used to define the end of the coda waves, as the SNR decays in time due to damping and geometrical attenuation.

\begin{figure}[!ht] \centering
	\includegraphics[scale=0.325]{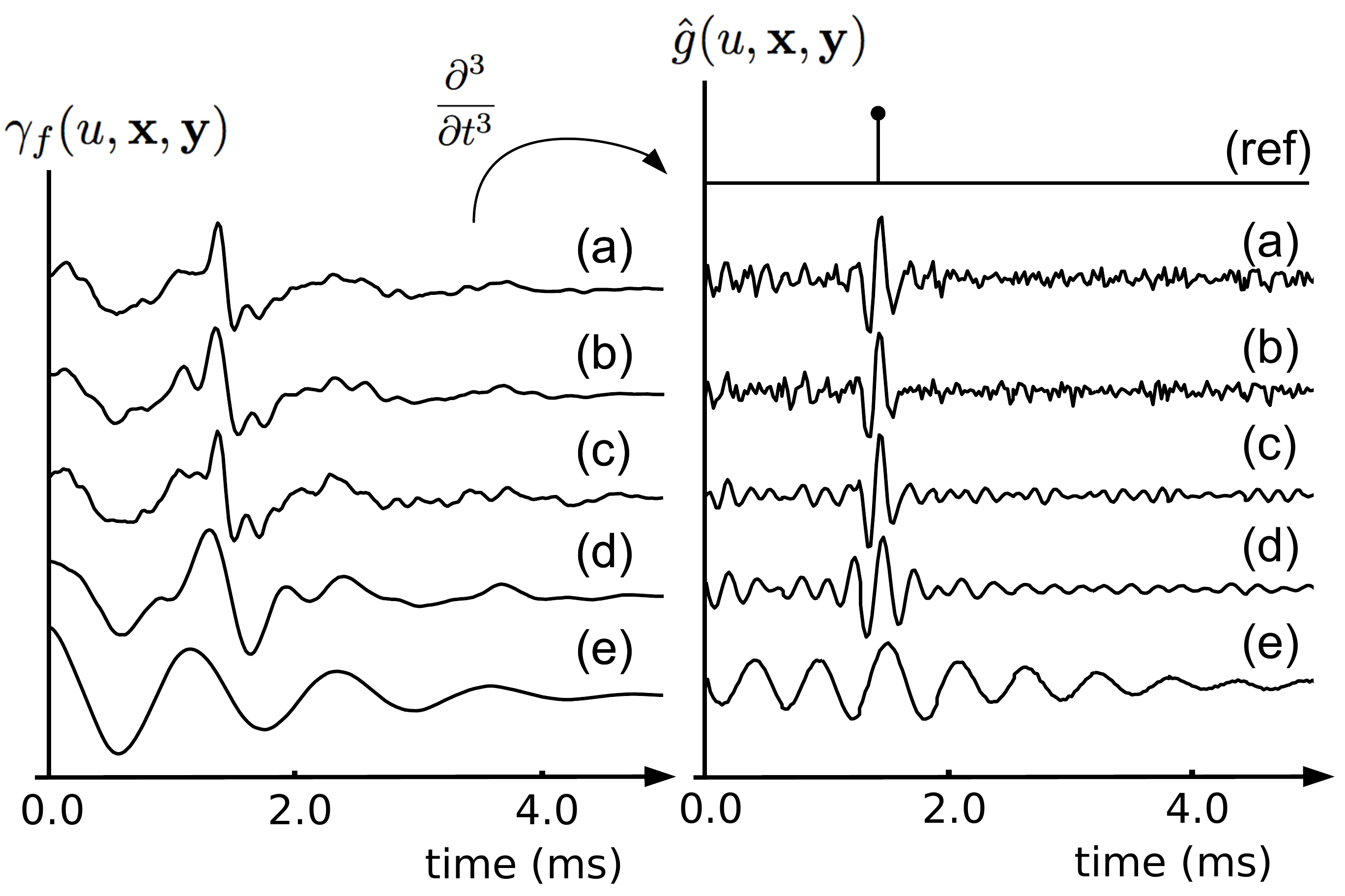}
	\caption{(left) Green's correlation estimated from noise correlation. Source fields were lowpass filtered, with cut-off frequencies $22$ kHz (a), $18$ kHz (b), $9$ kHz (c), $5$ kHz (d) and $2$ kHz(e). (right) Estimated Green's functions. The (ref) function locates the propagation delay at 1.40 ms. No equivalent reference for the Green's correlation is available in this experiment.}
	\label{f:correst}
\end{figure}

The Green's correlation was estimated from inter-correlations of coda waves, using the protocol described in Appendix \ref{a:protocol}. Extracted codas were lowpass filtered with various cut-off frequencies. For each bandwidth, 250 inter-correlations of 5-ms segments of the coda waves were averaged. Segments overlapped with each others and were smoothed by a Blackman apodization window. The Green's functions were estimated from the retrieved Green's correlations, by the application of the Ward identity \eqref{m:wig}. Results are displayed in Fig. \ref{f:correst}, in which the reference (ref) indicates the expected position of the propagation delay.

%
%
%
%
%

\begin{figure}[!ht] \centering
	\includegraphics[scale=.55]{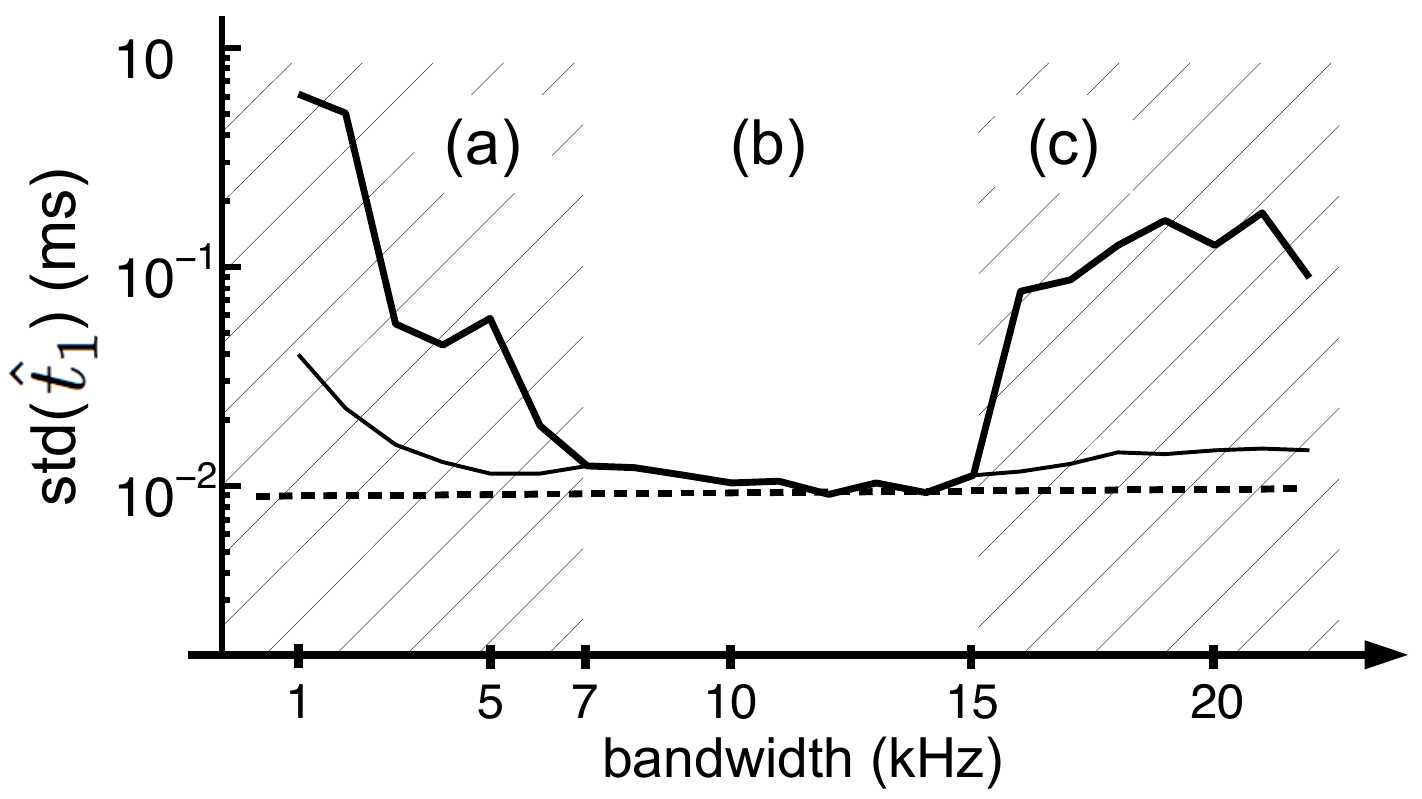}
	\caption{Mean standard deviation of the estimated propagation delay $\hat t_1$. (dotted line) Resolution due to the sampling frequency and the finite number of iterations, at $9.7\times 10^{-3}$ ms. (thick line) Performances computed from all propagation delays that were estimated. (fine line) Performances computed excluding misdetections. Discrepancies in the hatched zones are clear consequences of misdetection.}
	\label{f:std_perf}
\end{figure}

In this experiment, handclaps were found to have energy up to $15$ kHz. When lowpass filtering the signals, the SNR is thus expected to increase with the bandwidth until this threshold of $15$ kHz is reached, after which only measurement noise remains. The performances for the passive propagation delay estimation are thus expected to improve up until the cut-off frequency reaches 15 kHz.

In region (a), increasing the bandwidth leads to a decreasing number of misdetections, which in turn results in a smaller mean standard deviation. This result was predictable from Equation \eqref{m:perr}, that quantifies the probability of misdetection as a monotonic function of the SNR. As expected, the SNR increases with the bandwidth.

\begin{figure}[!ht] \centering
	\includegraphics[scale=.6]{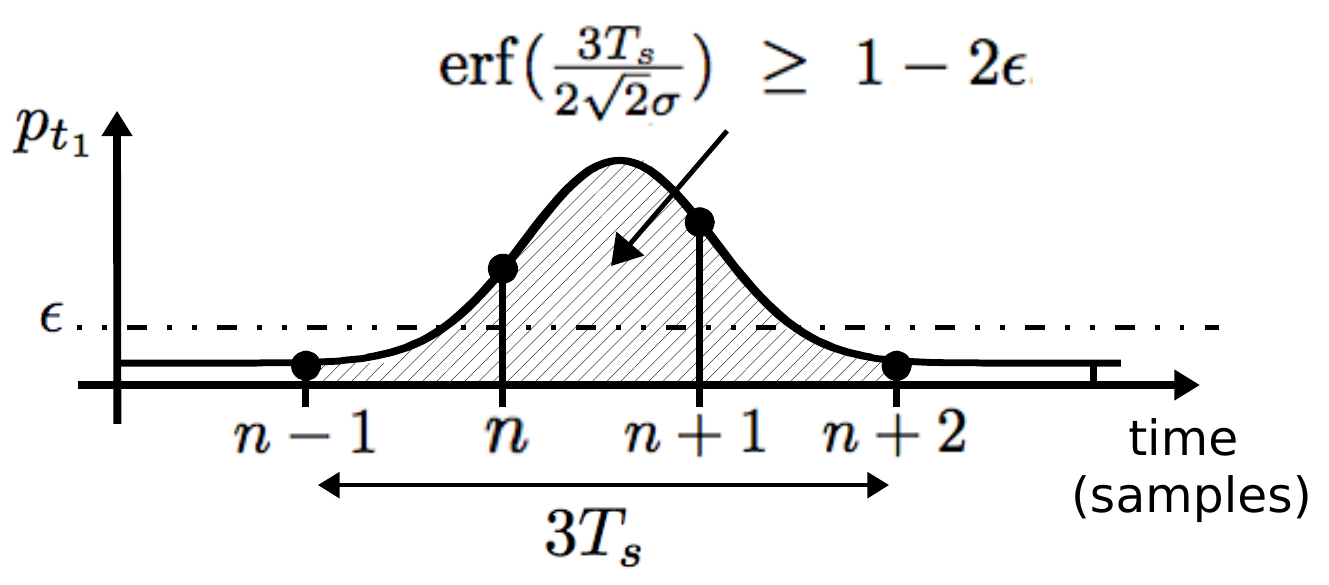}
	\caption{Distribution of the estimated propagation delay in aerial acoustics. Two samples are estimated with strong probabilities, resulting in a relationship between the variance of the estimator and the sampling rate.}
	\label{f:histtau}
\end{figure}

In region (b) (the cut-off frequency is in the range $[7-15]$ kHz), the probability of misdetection as defined in \eqref{m:perr} decreases below 0.001. The lower bound that was derived in Equation \eqref{m:pcrlb1} is overruled by the technical limitations of the experimental setup. More precisely, two factors come into play: the resolution of the estimation, due to the fact that the estimated propagation delays are multiples of the sampling period $T_s$, and the fact that the statistics are estimated over a finite number of iterations (1000 in this experiment). Fig. \ref{f:histtau} illustrates that mostly two neighboring locations are estimated with high probabilities. In other words, estimating any other locations than the sample $n$ or $n+1$ occurs with a probability that is set by a given threshold $\epsilon$. Then the probability that the propagation delay is contained inside a segment of length $3T_s$ is greater than 1-2$\epsilon$. Furthermore, if the probability distribution function $p_{t_1}$ of the propagation delay is modeled by a normal distribution with mean $t_1$ and standard deviation $\sigma$, the following relationship holds:~$\mathrm{erf}(\frac{3T_s}{2\sqrt{2}\sigma})\geq 1-2\epsilon$. In the experimental setup, $T_s=1/44100$ and $\epsilon = 10^{-3}$. The standard deviation thus meets a resolution bound at $9.7\times 10^{-3}$ ms, which is validated by the experimental results. For comparison, application of Equation \eqref{m:pcrlb1} using the data in Fig. \ref{f:correst} (c) to estimate the estimation noise level over the bandwidth leads to a theoretical bound for the standard deviation of $10^{-6}$ ms at $9$ kHz.

In region (c) (the bandwidth is larger than 15~kHz), misdetections become more likely, leading to an increase of the mean standard deviation. This is interpreted as a severe decrease of the SNR for bandwidths that are larger than 15~kHz, since handclaps have no significant energy above this frequency and only measurement noise is recorded.

As a summary, detecting the propagation delay is more likely to fail when the cut-off frequency is outside of the band $[7 - 15]$ kHz, which directly results in a strong increase of the standard deviation of the propagation delay estimation. On the opposite, when the cut-off frequency is inside of the band $[7 - 15]$ kHz, the experimental standard deviation is seemingly bounded, although it is only due of the finite sampling rate and the finite number of iterations of the estimation. Nevertheless, in the case of aerial acoustics, this limitation allows to retrieve inter-sensor distances with a standard deviation that is comparable to the dimensions of the microphones ($\approx1$~cm).
\section{An interpretation of the spatial distribution of the source field} \label{s:spatial}

In the previous experiment, coda waves were used to perform inter-sensor propagation delay estimation in a passive context. In this section, a model for coda waves is briefly presented via a geometrical analogy, in order to easily account for the spatial distribution of the source field. Dissipation is not accounted for. We show how the spatial distribution impacts the passive propagation delay estimation. The study is illustrated with simulations staging space-filtered sources.

\subsection{Towards a CRLB in the passive context} \label{ss:model}

In this analogy, the propagation medium is assumed boundary free. We show that any field that induces time differences of arrival at the sensor set could be replaced by a set of plane waves with particular incidence angles, as suggested in \cite{Sanchez}. This allows to model sources of opportunity (such as the coda wave) with a strong highlight on their spatial characteristics.

Consider a point-wise source at location $\vec{s}$ in the medium, with spherical wavefront. The sensors receive the field at different times, delayed by the time difference of arrival $\Delta$ that is due to the source-sensors geometry. It is  shown in ~\ref{a:geo} that for any point-wise source, there exists an injective transformation $\mc{F}$ that maps the source location $\vec{s}$ to an incidence angle $\theta_{eq}$, so that a plane wave impinging on the sensor set given this incidence angle would preserve the time difference of arrival,
\begin{equation}
\mc{F} :
\vec{s} \mapsto \theta_{eq} \textrm{, s.t. } \Delta(\vec{s}) = \Delta(\theta_{eq}).
\label{m:transF}
\end{equation}

\begin{figure}[ht!]
   	 \centering
	 \includegraphics[scale=0.8]{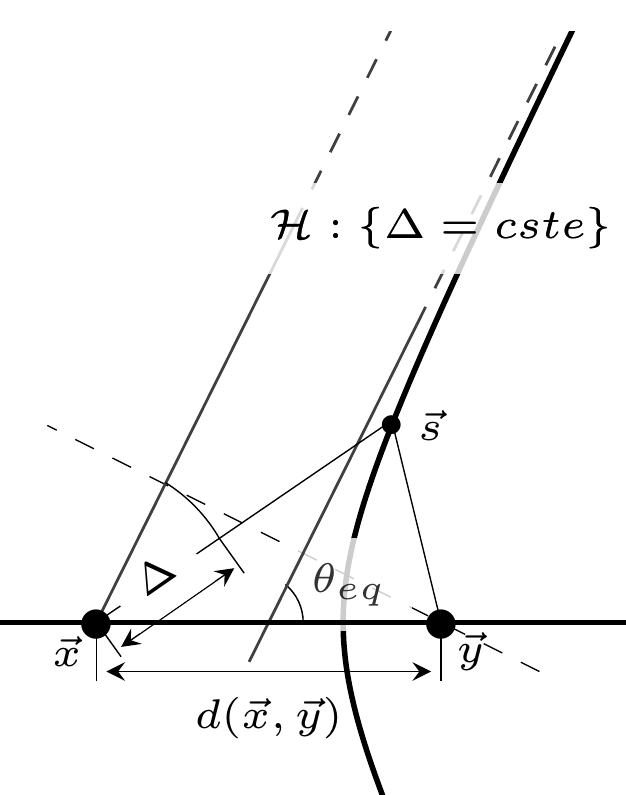}
	 \caption{Boundary-free medium analogy. For any time difference of arrival $\Delta$ induced by a point-wise source, there exists an incidence angle $\theta_{eq}$ such that a plane wave impinging on the sensor set preserves this delay.}
	 \label{f:anglearrival}
\end{figure}

Combining this geometrical property with the Huygens-Fresnel principle, any random source field can be modeled by an equivalent set of plane wave sources. Furthermore, in this analogy, a coda wave is thus replaced by a set of plane wave sources with a given distribution of incidence angles $p_\theta$, see Fig. \ref{f:anglearrival}.

To go further, consider a set of sources of opportunity that are i.i.d. in time and space. All fields thus have the same time correlation function $\gamma_s(t)$ and impinge on the set of sensors with incidences angles $\theta$ that have the same distribution $p_\theta$. Denote $f(t,\mb{x} | \theta_i)$ the field measured at position $\mb{x}$, due to the source field impinging with the incidence angle $\theta_i$ and amplitude $s(t,\theta_i)$. The inter-correlation of the fields measured at $\mb{x}$ and $\mb{y}$, depends on the associated time difference of arrival $\Delta_i$, illustrated in Fig. \ref{f:anglearrival}. In this analogy, the choice is made not to account for attenuation and to focus only on the distribution of the sources, in which case the correlation of the fields is
\begin{equation}
\mbb{E}\big[ f(t,\mb{x} | \theta_i) f^*(t+u,\mb{y} | \theta_j)  \big] = \bar\gamma_s(u-\Delta_i) \delta(\theta_i - \theta_j).
\label{m:1srctdoa}
\end{equation}
where $\bar\gamma_s$ is normalized correlation of the source.

When using sources of opportunity, the field amplitudes of interest, \textit{e.g.} coda waves, are random processes whose statistics allow to estimate the Green's correlation of the medium. Everything else is treated as a perturbation, \textit{e.g.} a measurement noise. Therefore, property \eqref{m:1srctdoa} is fundamental to derive an observation model for the Green's correlation function. Combined with the correlator in  \ref{a:protocol}, the Green's correlation is estimated by
\begin{equation}
\hat\gamma_{g}(t) = \int_0^{2\pi} p_\theta(\theta) \gamma_s\big( t - {t_1}\cos(\theta)\big) d\theta + e(t).
\label{m:rgf}
\end{equation}
where $\gamma_s(t)$ is the auto-correlation function of the source. It is estimated from signals that contain measurements noises and that have finite-length, see  \ref{a:protocol}. Corresponding estimation errors are represented by the additive noise $e(t)$. For the sake of simplicity, $e(t)$ is approximated centered Gaussian, see \cite{VanTrees} for some justifications.

Equation \eqref{m:rgf} provides a novel observation model, directly on the Green's correlation function. The Fisher information on the minimum propagation delay ${t_1}$ writes
\begin{equation*}
F_{t_1}(p_\theta) = \frac{1}{\sigma^2} \int_\mbb{R} \Big[ \int_0^{2\pi} p_\theta(\theta) \cos(\theta) \frac{\partial \gamma_s}{\partial t} ( t - {t_1}\cos(\theta)) d\theta \Big]^2 dt.
\label{m:fisher2}
\end{equation*}

%

As in \cite{VanTrees}, a Cramer-Rao Lower Bound for the passive context can thereby be derived as
\begin{equation}
\mbb{E}\big[ (\hat {t}_1 - {t_1})^2 \big] \geq \frac{1}{F_{t_1}(p_\theta)}.
\label{m:msebound1}
\end{equation}

In order to conduct the calculus in Equation \eqref{m:wig}, the strong assumption of spatial whiteness of the source field was assumed. However, spatial properties of the sources are expected to impact the performances of passive estimators, just as temporal properties do. 
In the following, the new observation model \eqref{m:rgf} is used to run simulations and the passive Cramer-Rao Lower Bound \eqref{m:msebound1} is computed and compared to the obtained results.

\subsection{Illustrations with space-filtered sources} \label{ss:space_filtered_sources}

From the model introduced in section \ref{ss:model}, four setups are investigated in order to study typical and realistic pathologies for the spatial distribution of the source field. (1) The first setup involves uniformity of the source field; (2) the second setup stages setup 1 with a small count of sources; (3) the third setup stages setup 1 with reduced spatial bandwidth (solid angle containing the impinging wave vectors). (4) Finally, the fourth setup investigates what happens when the source field is not uniformly distributed. Eventually, performances are compared to the theoretical bounds obtained in this paper. 

\subsubsection{Simulations}

\begin{figure}[ht!] \centering	
	\includegraphics[scale=0.4]{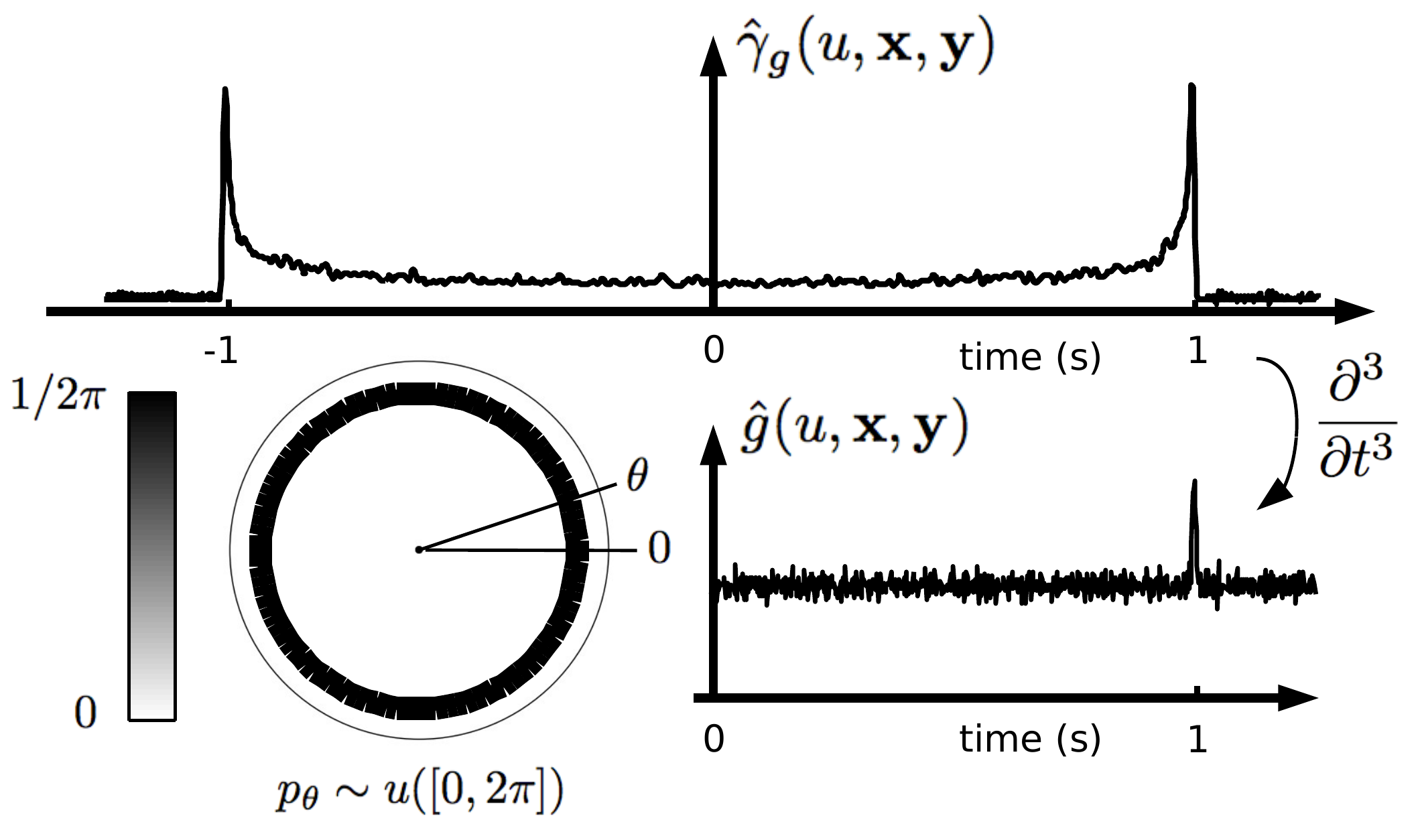}
	\caption{Setup (1). Uniform source field scenario. Number of sources $N = 10^5$ and bandwidth $L$. (bottom left) Incidence angles are uniformly distributed over the whole phase space. (bottom right) Provided that $NL \gg 2t_1$, the differentiation of the Green's correlation leads to the presence of a peak at $1$ s. This is well matched with model \eqref{m:gmodel} for a boundary free medium.}
	\label{f:unifcv}
\end{figure}

\begin{figure}[ht!] \centering
	\includegraphics[scale=0.4]{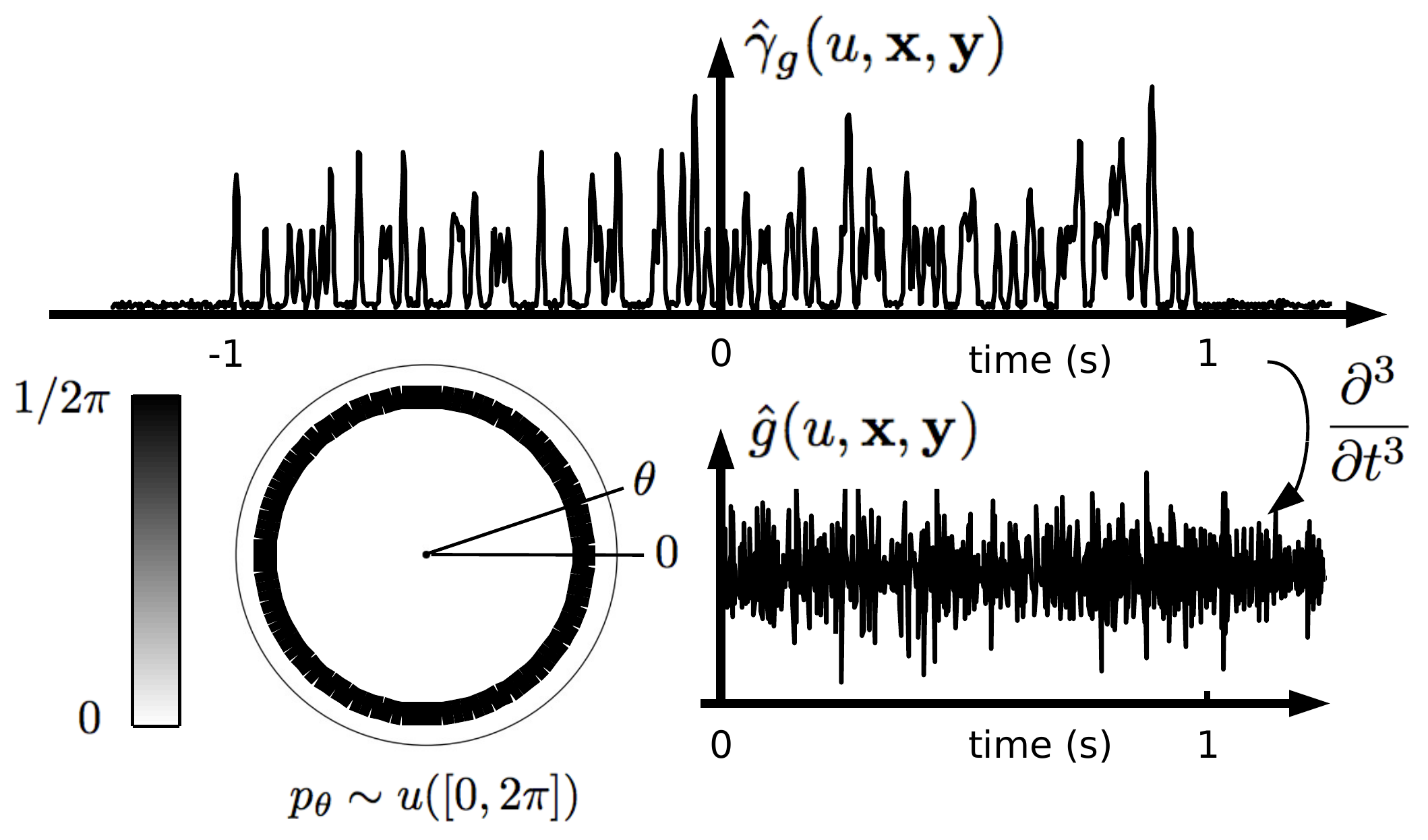}
	\caption{Setup (2). Small number of sources scenario. Number of sources: $10^2$ and bandwidth $L$. (bottom left) Incidence angles are uniformly distributed over the whole phase space. (top) When $NL \leq 2t_1$, the Green's correlation is not properly estimated. (bottom right) As a consequence, the differentiation of the Green's correlation is very inconsistent with model \eqref{m:gmodel} and thus denies propagation delay estimation.}
	\label{f:unifnocv}
\end{figure}

\begin{figure}[!ht] \centering
	\includegraphics[scale=0.4]{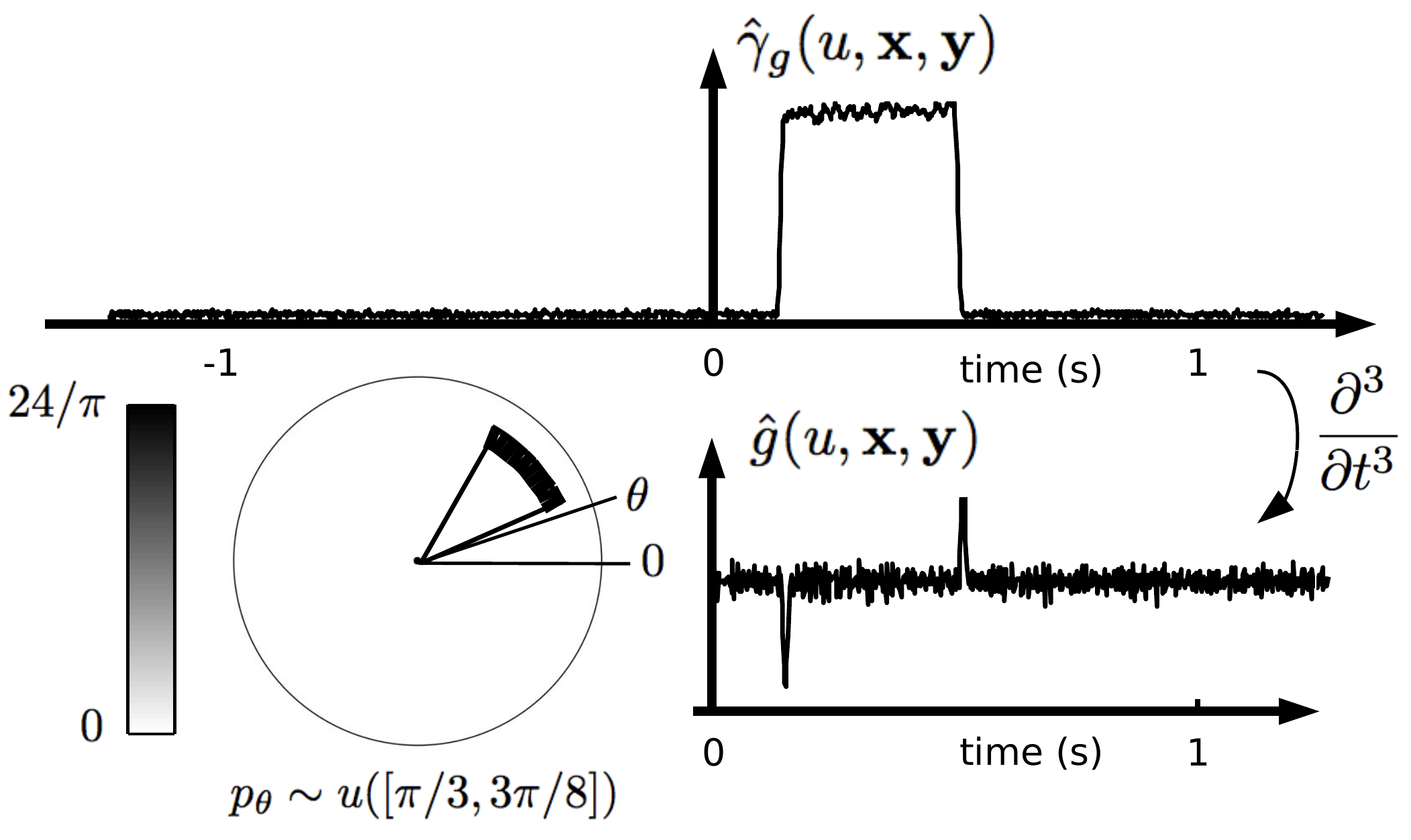} 
	\caption{Setup (3). Band-limited source field scenario. Number of sources: $10^5$ and bandwidth $L$. (bottom left) As a special case from scenario in Fig. \ref{f:unifcv}, incidence angles are uniformly distributed in a small subset of the phase space. This kind of setup is similar to that observed by \cite{Derode} and \cite{Prieto} in a basin in California where the impinging directions of waves were confined to a subset of the whole phase space. (bottom right) Here, a bias $B(\hat{t_1}) = {t_1}/2$ is created in the propagation delay estimation. Note however that if $\theta = 0$ is contained in the distribution, the estimation of the propagation delay in this setup may still be achievable.}
	\label{f:uniffilt}
\end{figure}

\begin{figure}[!ht] \centering
	\includegraphics[scale=0.4]{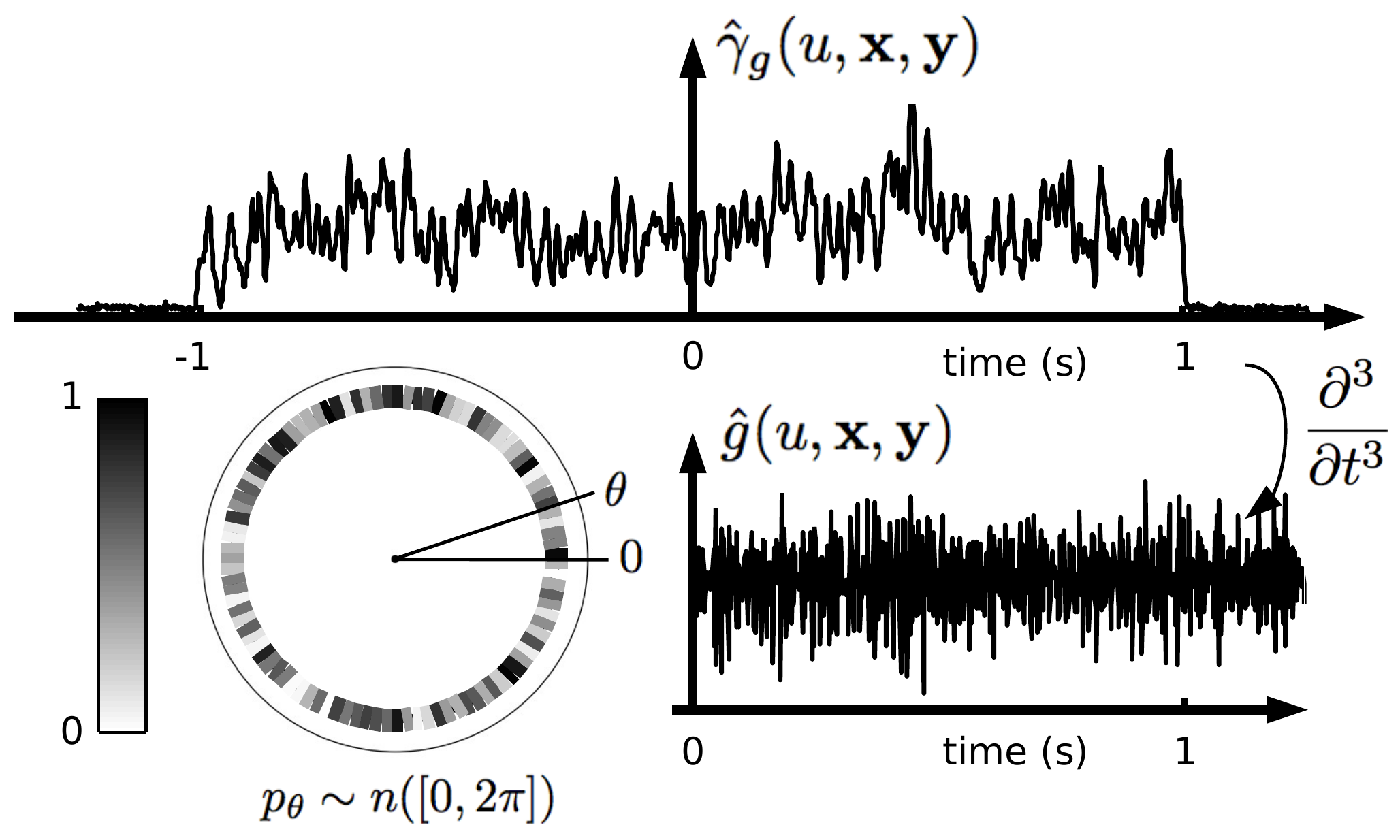} 
	\caption{Setup (4). Non uniform source field scenario. Number of sources: $10^5$ and bandwidth $L$. (bottom left) Incidence angles are non uniformly distributed over the whole phase space. (top) The function that is estimated in place of the Green's correlation can be seen as a space-filtered version of it. Thus, the estimated Green's function prevents parameters extraction from the application of estimators introduced for time-filtered sources only, such as the one in Equation \eqref{m:test}.}
	\label{f:nonunif}
\end{figure}

Time differences of arrival are confined to $[-1,1]$ s, with the inter-sensor distance set to $1$~m and the local propagation speed of the wave equal to $1$~m/s. The propagation delay $t_1$ to estimate is thus located at $1$~s. A set of i.i.d. plane wave sources impinges on the sensor set. The auto-correlation function of a source is the sinc function with a large bandwidth. As presented in Equation \eqref{m:cgest}, each correlation is normalized by its power. Each correlation is then degraded by some additive white sequence of Gaussian noise with variance set to $0.01$. In Fig. \ref{f:unifcv} -- \ref{f:nonunif}, the top quadrant displays the estimated Green's correlation. The distribution of the incidence angles that leads to this estimation of the Green's function is displayed in the bottom-left. The color bar indicates the intensity map for the values of the distribution $p_\theta$. The estimated Green's function is displayed on the bottom right.

Note at this point that an alternative estimator for the propagation delay can be formed by retaining the largest time difference of arrivals. This estimator is only valid in this setup which is the reason why it is not studied here.

From the conducted simulations, uniformity of the distribution of incidence angles coupled to a large number of sources appears essential to achieve passive propagation delay estimation. A slight deviation from the uniform distribution leads to biased estimations. 

\subsubsection{Performances study}

Two observation models were derived for the purpose of propagation delay estimation. The model in Equation \eqref{m:gmodel2} is a band-limited pulse in some noise, which is applicable to the estimated Green's function --- whether in an active or a passive context. The model in Equation \eqref{m:rgf} is specific to the passive approach as it focuses on the Green's correlation function. From these models, two theoretical lower bounds for the variance of the propagation delay estimation were derived, in Equations \eqref{m:msebound1} and \eqref{m:pcrlb1}.

\begin{figure}[!ht] \centering
	\includegraphics[scale=0.5]{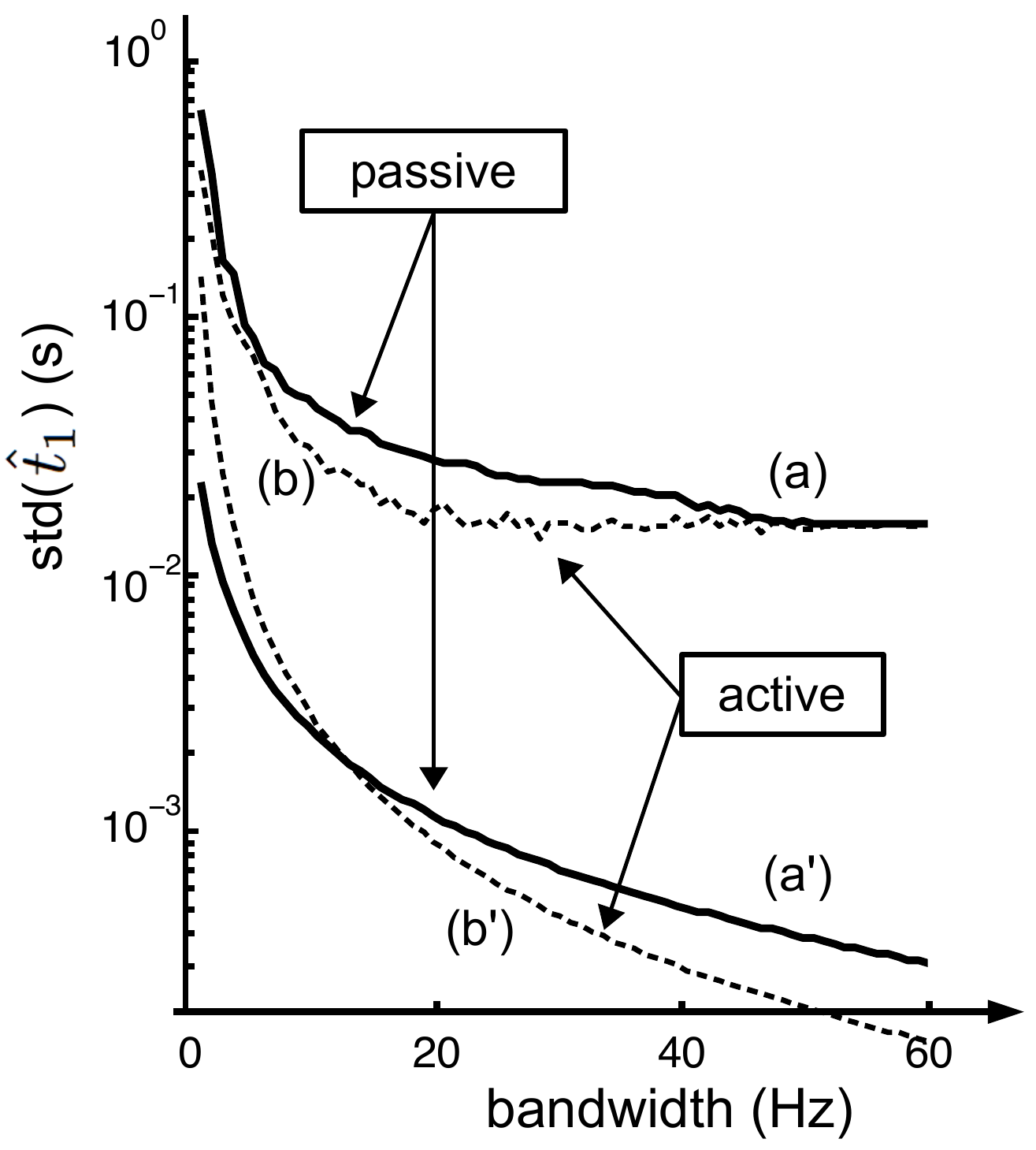} 
	\caption{Standard deviation for the propagation delay estimation. (thick lines) (a) Performances for the passive propagation delay estimation (a') Theoretical lower bound in the passive case, from the model of the Green's correlation, in Equation \eqref{m:msebound1}. (dotted lines) (b) Performances obtained in the active case, based on model \eqref{m:gmodel2}. (b') Theoretical lower bound from Equation \eqref{m:pcrlb1}, applicable for both the passive and the active case.}
	\label{f:pCRLB}
\end{figure}

The passive propagation delay estimation performances are computed based on setup (1), with a source field that has a varying bandwidth. The mean standard deviation for the passive and the active estimation of the propagation delay are compared to the theoretical bounds, see Fig. \ref{f:pCRLB}.

The theoretical CRLB in the passive context is 2 decades lower than the actual standard deviation that was obtained in simulation. This discrepancy was expected, since the bound \eqref{m:msebound1} was obtained from the observation model \eqref{m:rgf}, which describes the Green's correlation function. The propagation delay was not directly estimated from this data. Instead, the Green's function was first retrieved by applying Equation \eqref{m:wig} and only then was the propagation delay extracted. Thus, as quantified in  \ref{a:protocol}, huge performances losses were expected. The performances in the active context are sensibly better than in the passive context but both reach an experimental lower bound as in the real-data experiment presented in section \ref{ss:exp}.
\section{Perspectives and conclusions}

In this article, identification of a linear wave propagation medium was introduced in a passive context, \textit{i.e.} using sources of opportunity only. The ward identity was extended to account for the spectral properties of acoustic sources. Estimators for the Green's correlation function, the Green's function and an example of parameter, the propagation delay, were then studied in the light of the temporal statistics of the source field. Performances from real-data experiments and simulations were compared to the theoretical bounds introduced in this paper.

In the studied experimental setup, detection errors for the passive propagation delay estimation were shown to be less likely when the acoustic signals are lowpass filtered below 7kHz to 10kHz. In this case, performances were bounded by the temporal resolution due to the finite sampling rate. The accuracy of the inter-sensor distance estimation is comparable to the size of the sensors ($\approx 1$~cm).

Based on a geometrical analogy, an observation model with plane waves was derived for the passive context, allowing to account for the spatial properties of the source. It was shown that non uniformity of the propagation directions of the field, over the whole phase space, generally leads to severe estimation errors. This analogy leads to an observation model on the Green's correlation function and the derivation of the accompanying Cramer-Rao lower bound. The passive and active estimation protocol were found to have comparable performances.

Finally, it seems very beneficial to develop a technique that would perform passive propagation delay estimation from the Green's correlation function, since the theoretical bound is attractively lower than the obtained experimental results. Furthermore, a model-based approach would allow to estimate parameters with a resolution that would rely less on the sampling rate, which was the limiting factor for the performances in the studied experiments.

An extension with great potential in Structural Health Monitoring includes the passive localization of embedded or burrowed sensors and starts by extending the Ward identity to the concerned medium (\textit{e.g.} concrete).

\appendix

\section{Passive identification protocol} \label{a:protocol} 

Measurements at locations $\vec{x}$ and $\vec{y}$ are times series $\mb{x}[n]$ and $\mb{y}[n]$. Acquisition is assumed sampled at frequency $1/T$ Hz. Quantities of interest are the pressure amplitudes due to some sources of opportunity, that are seen by both sensors. Measurement noises are assumed to be i.i.d. zero-mean Gaussian random sequences, $\mb{w}_x(t)$ and $\mb{w}_y(t)$ respectively. The observation model is thus 
\begin{equation*}
\left \{
\begin{array}{clc}
\mb{x}[n] &=& f(nT , \mb{x}) + \mb{w}_x[n] \\
\mb{y}[n] &=& f(nT , \mb{y}) + \mb{w}_y[n]
\end{array}
\right. .
\label{m:data_model}
\end{equation*}

In practice, the Green's correlation $\gamma_{g}(t)$ in Equation \eqref{m:cg} is estimated from finite-length and noisy signals. 
Expectations over time and space can be computed by splitting the signals $\mb{x}[n]$ and $\mb{y}[n]$ into $K$ segments using a sliding window of constant duration. The average of all windows allows to estimate the spatial expectation, seen as an ensemble average. The $k$th window is thus the measurement between the starting index $a_k$ and the ending index $b_k$ where $|b_k - a_k|$ is constant for all $k$. Note that overlap between windows is possible. An estimation of the Green's correlation, can be
\begin{equation}
\hat\gamma_{g}[n] = \frac{1}{K} \sum_{k=1}^K \frac{1}{(b_k-a_k)\sqrt{E_k}} \sum_{l=a_k}^{b_k-n} \mb{x}[l] \mb{y}[l-n].
\label{m:cgest}
\end{equation}

In Equation \eqref{m:cgest}, the correlation estimator is biased but has low variance \cite{VanTrees}. The bias is convolutive (Bartlett window) but does not impact the position of the max. The emphasis is then put on the variance which has to be kept as low as possible. The normalizing constant $E_k = \sum_k x[k]^2 \sum_k y[k]^2 $ is the power associated to the fields in the $k$th window. This normalization is important since inequalities on the energies of the fields can be due to inequal power of sources, geometrical attenuation and dissipation, but do not necessarily imply uncorrelatedness of the fields.

By the application of the Ward identity \eqref{m:wi_wviscous}, the Green's function is retrieved by differentiating the correlation obtained in Equation \eqref{m:cgest}. Differentiation can be numerically computed using different techniques. Whatever the technique, it is known that the variance of signal differentiated greatly increases. To provide an order of magnitude, consider the basic case of differentiating using subtractions. Let $\mb{x}[n]$ be some zero-mean i.i.d. Gaussian random sequence. Its variance is $\sigma^2$. The $\alpha$-th differentiation of $\mb{x}[n]$ can be numerically computed using the formula $\mb{x}[n]^{(\alpha)} = \mb{x}[n]^{(\alpha-1)} - \mb{x}[{n-1}]^{(\alpha-1)}$. A simple proof leads to the conclusion that the variance of the $\alpha$th order differentiation of $\mb{x}[n]$ is $\binom{2\alpha}{\alpha} \sigma^2$, where $\binom{a}{b}$ is the binomial coefficient. The order of the differentiation is given by the dissipation model. In the case of weakly viscous damping, $\alpha = 3$, see Equation \eqref{m:wi_wviscous}. Consequently, the variance of the estimation residue of the Green's function is 20 times the variance of the estimation residue in the Green's correlation. In comparison, the constant damping model used in geosciences leads to a factor $2$ instead. As a consequence, although differentiation is a necessary step in the passive estimation protocol, if one wants to retrieve the Green's function, one has to bear in mind that the variance of the estimation error will always increase, and that it depends on the damping model.

\section{A geometrical analogy} \label{a:geo} 

The relationship between the time difference of arrival and the inter-sensor propagation is a function of the source position, see Fig. \ref{f:anglearrival}. In a passive context, the position of the source is uncontrolled. We would like to link the time difference of arrival $\Delta$ that is induced by any source to an equivalent incidence angle $\theta_{eq}$ of a plane wave source that would preserve exactly the same time difference of arrival
\begin{equation*}
\Delta(\vec{s})  =   \Delta (\theta_{eq} )   =   t_1 \cos(\theta_{eq}),
\end{equation*}

The position of the source is allowed to change while still preserving the time difference of arrival
\begin{equation*}
\forall \vec{s}, \quad | ~d(\vec{s},\vec{x}) - d(\vec{s},\vec{y})~ |   =   v\Delta.
\end{equation*}
where $v$ is the wave velocity and $d(\vec{a},\vec{b})$ is the Euclidean distance between locations $\vec a$ and $\vec b$. This relationship naturally casts the set of the positions allowed into a hyperbola $\mc{H}$ with the reduced equation
\begin{equation*}
\frac{x^2}{a^2} - \frac{y^2}{b^2}  =   1.
\end{equation*}

The hyperbolas intersect the $[\mb{x}; \mb{y}]$ segment at $a = \Delta/2$, which then allows to find $b  =   \frac{1}{2}\sqrt{ (vt_1 + \Delta)(vt_1 - \Delta ) }$. Eventually, recall that hyperbolas have asymptotes with slopes $\alpha = \pm b/a$, which defines the incidence angle of the equivalent plane-wave source
\begin{equation*}
\theta_{eq}  =  \arctan\big(\frac{1}{ 2\Delta }\sqrt{ (vt_1 + \Delta)(vt_1 - \Delta ) }\big).
\end{equation*}

As a conclusion, any time difference of arrival can be infered by a unique equivalent plane-wave source.


\section*{References}
\begin{thebibliography}{00}

\bibitem{Aki} K. Aki and B. Chouet, {Origin of coda waves: source, attenuation and scatterin effects}, J. Geophys. Res., vol. 80, pp. 3322--3342 (1975).
\bibitem{Blandin} C. Blandin, A. Ozerov, E. Vincent, {Multi-source TDOA estimation in reverberant audio using angular spectra and clustering}, Signal Processing, Volume 92, Issue 8, pp. 1950-1960 (2012).
\bibitem{Buckingham} M.J. Buckingham, {Causality, Stokes' wave equation and acoustic pulse propagation in a viscous fluid}, Physical Letter E 72 (2005).
\bibitem{Campillo} M. Campillo, A. Paul, {Long-range correlations in the diffuse seismic coda}, Science 299 (2003) 547-549.
\bibitem{Campillo2} M. Campillo, Phase and correlation in "random" seismic fields and the reconstruction of the Green function, Pure and Applied Geophysics 163 (2006) 475-502.
\bibitem{Carmona} M. Carmona, O. Michel, J-L. Lacoume, B. Nicolas, N. Sprynski,  {Ward identities for visco-acoustic and visco-elastic propagation media}, Wave motion 49 (2012) 484-489.
\bibitem{Chalise} B. Chalise, Y. Zhang, M. Amin, B. Himed, {Target localization in a multi-static passive radar system through convex optimization}, Signal Processing, Volume 102, pp. 207-215 (2014).
\bibitem{ColinDeVerdiere} Y. Colin de Verdiere, {Semi-classical analysis and passive imaging}, NonLinearity 22 (2009) 45-75.
\bibitem{Derode} A. Derode. E. Larose, M. Campillo, M. Fink, {How to Estimate the Green's Function of a Heterogeneous Medium Between Two Passive Sensors ? Application to Acoustic Waves}, Applied Physics Letters 83 (2003) 3054-3056.
\bibitem{Dumont} T. Dumont, S. Le Corff, {Simultaneous localization and mapping in wireless sensor networks}, Signal Processing, Volume 101, pp. 192-203 (2014).
\bibitem{Farrar} C.R. Farrar, G.H. James III, System Identification from Ambient Vibration Measurements on a Bridge, Journal Sound and Vibration 1 (1999).
\bibitem{Gouedard} P. Gou\'edard, L. Stehly, F. Brenguier, M. Campillo, Y. Colin de Verdi\`ere,E. Larose, L. Margerin, P. Roux, F.J. S\`anchez-Sesma, N.M. Shapiro, R. Weaver, Cross-correlation of random fields: mathematical approach and applications, Geophysical Prospecting 56 (2008) 375-393.
\bibitem{Kay1993} Steven M. Kay, {Fundamentals of Statistical Signal Processing : Estimation Theory}, Prentice Hall signal processing series (1993). Chapter 3.11.
\bibitem{Kubo} R. Kubo, {The fluctuation-dissipation theorem}, Reports on Progress in Physics, 29, 255 (1966).
\bibitem{Lacoume} J. L. Lacoume, {Tomographie passive, observer avec du bruit}, GRETSI (2007). 
\bibitem{Leung} H. Leung, T. Lo, J. Litva, {Angle-of-arrival in multi path environment using chaos theory}, Signal Processing, Volume 31, pp. 57-68 (1993).
\bibitem{Lobkis} O. Lobkis, R. Weaver, On the emergence of the Green function in the correlations of a diffuse field, J. Acoust. Soc. Am. 110 (2001) 3011-3017.
\bibitem{Nyquist} H. Nyquist, {Thermal agitation of electric charges in conductors}, Phys. Rev. 32, 110 (1928).
\bibitem{Olabode} O. Olabode, P. Enikanselu, {Analysis of seismic time-depth conversion using geostatistically-derived averaged velocities over "Labod" field, Niger Delta, Nigeria.}, Ozean Journal of Applied Sciences 1 (1), (2008).
\bibitem{Patwari} N. Patwari, {Location estimation in sensor networks}, PhD in University of Michigan (2005).
\bibitem{Prieto} G. A. Prieto, {Anelastic Earth structure from the coherency of the ambient seismic field}, Journal of Geophysical Research, vol. 114, B07303, 2009.
\bibitem{Roux} P. Roux, K.G Sabra, W.A. Kuperman, A. Roux, {Ambient noise cross correlation in free space: Theoretical approach}, J. Acoust. Soc. Am. 117 (2005) 79-84.
\bibitem{Sabra} K.G. Sabra, P. Roux, A.M. Thode, G.L. D'Spain, W.S. Hodgkiss, W.A. Kuperman, Extracting time-domain Green function estimates from ambient seismic noise, IEEE J Oceanic. Eng. 30 (2005) 338-347.
\bibitem{Sanchez} F. Sanchez-Sesma, {Retrieval of the Green function from cross-correlation: the canonical elastic problem}, Bulletin of Seismological Society of America (2006).
\bibitem{Schuster} A. Schuster, {An introduction to the Theory of Optics}, E. Arnold (1904).
\bibitem{Siringoringo} D. Siringoringo, Y. Fujino, {System identification of suspension bridge from ambient vibration response}, Engineering Structures 30 (2008), pp. 462-477.
\bibitem{Snieder} R. Snieder, K. Wapenaar, U. Wegler, Unified Green function retrieval by cross-correlation; connection with energy principles, Phys. Rev. E 75 (2008) 0361031-03610314.
\bibitem{Sthely} L. Sthely, N. Shapiro, {travel time measurements from noise correlation: stability and detection of instrumental time-shifts}, Geophysics. Int. J., 171 (2007) pp. 223-230.
\bibitem{Stokes} G.G. Stokes, Trans. Cambridge Philos. Soc. 8 287 (1845).
\bibitem{Vincent} R. Vincent, M. Carmona, O. Michel, {Passive acoustic sensor network localiation; application to structure geometry monitoring}, EWSHM (2014).
\bibitem{VanTrees} H. L. Van Trees, K. L. Bell, Z. Tian, {Detection, estimation and filtering theory}, Wiley, Second Edition 2013.
\bibitem{Wapenaar} K. Wapenaar, E. Slob, R. Snieder, and A. Curtis, {Tutorial on seismic interferometry. Part II: Underlying theory and new advances}, Geophysics 75 (2010) 75A211-75A227.
\bibitem{Weaver} R. Weaver, {Ward identities and the retrieval of Green's functions in the correlations of a diffuse field}, Wave Motion 45 (2008) 596-604.
\bibitem{Zhao} S. Zhao, T. Saluev, D. Jones, {Undetermined direction of arrival estimation using acoustic vector sensor}, Signal Processing, Volume 100, pp 160-168 (2014).
\bibitem{Ziv} J. Ziv, M. Zakai, {Some Lower Bounds on Signal Parameter Estimation}, IEEE Transactions on Information Theory 15, 1998.
\end{thebibliography}
\end{document}